\documentstyle[12pt]{article}
 1
 1
 1

\begin{document}

\centerline{\bf Nuclear Quadrupole Effects in Deeply Bound 
Pionic Atoms}
\baselineskip=16pt

\vspace*{0.6cm}
\centerline{ N. Nose-Togawa$^{a,b)}$, S. Hirenzaki$^{a)}$ and
K. Kume$^{a)}$}
\baselineskip=16pt
\centerline{\it a) Department of Physics, 
Nara Women's University, Nara 630, Japan}
\centerline{\it b) RCNP, Osaka University, Ibaraki, Osaka 567,
Japan}

\vspace*{0.9cm}
\noindent
{\bf Abstract} \\

\leftskip=1cm
\rightskip=1cm 

We have studied nuclear quadrupole deformation effects in 
deeply bound pionic atoms theoretically.  We have evaluated the level 
shifts and widths of the hyperfine components using the first order 
perturbation theory and compared them with the effects of neutron skin.  
We conclude that the nuclear quadrupole deformation effects for 
deeply bound $1s$ and $2p$ states are very difficult to observe 
and that the effects could be 
observed for $3d$ states. We also conclude that the deformation effects 
are sensitive to the parameters of the pion-nucleus optical potential. \\  


\vspace*{0.8cm}
\normalsize\baselineskip=15pt
\setcounter{footnote}{0}
\renewcommand{\thefootnote}{\alph{footnote}}
\leftskip=0cm
\rightskip=0cm 

\vspace{.5cm}
\noindent
{\bf 1. Introduction} \\

Deeply bound pionic atoms such as $1s$ and $2p$ states 
in heavy nuclei were predicted to be quasi-stable \cite{Friedman85}
and were expected to be observed by proper experimental methods 
\cite{Toki88,Toki89}.    
There have been a number of efforts  
to find these states both experimentally and theoretically 
[4-10].  Very recently, deeply bound pionic 
atoms have been discovered by Yamazaki 
$et$ $al.$ \cite{Yamazaki96} using (d,$^3$He) reactions 
\cite{Hirenzaki91,Toki96}.  In the data we clearly identify a peak structure in  
the energy spectrum of the $^3$He which is mainly due to the formation of deeply  
bound $2p$ atomic state with a $2p^{-1}_{3/2}$ neutron hole state. 
Formation of the deepest $1s$ atomic state is also studied theoretically 
\cite{Hirenzaki96} and will be tried to find experimentally in near future 
\cite{Hayano96}.  
We can observe the energy levels and widths of the atomic states by 
knowing the reaction $Q$-value of the peak position 
and the peak width of the $^3$He spectrum.  Hence, we 
can expect to develop precise spectroscopic studies of deeply bound 
pionic atoms using 
the reaction with high energy-resolution.  
Since deeply bound pionic atoms exist very close to nuclear surface 
\cite{Toki88,Toki89}, and 
since they can be populated in all heavy nuclei including $\beta$ 
unstable ones using the (d,$^3$He) reactions in principle \cite{Yamazaki91}, 
we can expect to obtain new information on the nuclear surface 
and pion-nucleus interaction
from the spectroscopic studies of the deeply bound pionic states.  

Nuclear quadrupole deformation effects on pionic processes have been 
studied both in bound states \cite{Scheck72,Olaniyi82} and scattering 
states \cite{Nose94}.  In the previous works for the shallow bound states 
\cite{Scheck72,Olaniyi82}, 
the quadrupole effects were studied in  order to extract information on the 
nuclear deformation using experimental pionic atom spectra.  
In this paper, 
we study the quadrupole effects for deeply bound pionic states 
from different point of view.  
In order to know the property of the deeply bound pionic atoms well 
and to consider how we can obtain new information from the pionic atoms, we need  to 
evaluate the nuclear deformation effects.  If the effects are large enough, 
we may be able to use them to observe nuclear deformation precisely, and if the  
effects are small enough, we can just forget them and extract other 
information like pion-nucleus interaction and/or neutron skin.  
Since the deeply bound 
states have been observed clearly only once, our knowledge is very restricted and 
we need to study their properties in order to develop the spectroscopic study of  
the deeply bound pionic states. 
So far the deformation effects were  
evaluated only for 
the shallow pionic atoms such as $3d$ and $4f$ in heavy 
nuclei and were known to cause level splitting and width change of 
the atomic states \cite{Scheck72,Olaniyi82}.  
We can expect to have larger 
deformation effects for deeper bound states since pion exists closer to 
nucleus and, thus, feels nuclear surface more sensitively.  

In this paper we evaluate the nuclear quadrupole deformation effects on 
the deeply bound pionic atoms to know how important the effects are 
in pionic atom spectroscopy.  We use the first 
order perturbation theory to evaluate the level shifts and widths 
for the individual hyperfine components, which is described 
in section 2.  In section 3 we show the numerical 
results.  We summarize this paper in section 4.  \\  

\noindent
{\bf 2. Quadrupole effects on pionic atoms}        \\

In this section we describe briefly our theoretical formalism which is 
used to evaluate the quadrupole deformation effects on deeply bound pionic 
atoms.  We apply the first 
order perturbation theory \cite{Scheck72}.  

The nuclear density is defined by the expectation value of the 
nuclear ground state with spin $J$ and its z-projection 
$J_z = J$ as,

\begin{equation}
\rho ( {\bf r}) = < J J \mid \hat \rho ({\bf r}) \mid J J > ,
\end{equation}

\noindent 
where $ \hat \rho ({\bf r}) $ is defined as, 

\begin{equation}
\hat \rho ({\bf r}) = \sum_{i=1}^A \delta ( {\bf r} - {\bf r}_i ) ,
\end{equation}

\noindent 
with nuclear mass number $A$.  By multipole expansion, we can rewrite the 
density as, 

\begin{equation}
\rho ( {\bf r}) =  \rho_0(r) + \sum_{{\rm even} \  k\geq 2 } \sqrt{\frac{2k+1}{1 6 \pi}} \rho_k (r) 
Y_k^0 ( {\hat{\bf r}}) , 
\end{equation}

\noindent 
where for $k \geq 2$ 

\begin{equation}
\rho_k (r) = \sqrt{\frac{16 \pi}{2k+1}} < J J \mid \sum_{i=1}^A 
\frac{\delta ( r - r_i) }{r^2} Y_k^0 ( {\hat{\bf r}}_i) \mid J J > .
\end{equation}

\noindent
Here, the $k=2$ term corresponds to the nuclear quadrupole 
density which we will consider in this paper.  
As a practical parametrization, 
we use the Woods-Saxon form for the 
nuclear density distribution in the intrinsic frame:

\begin{equation}
\bar{\rho}( {\bf r}) = \frac{\rho_N}{1+{\rm exp\/}( \frac{r - R(\theta)}{a})}  
\end{equation}

\noindent
with 

\begin{equation}
 R(\theta) = R ( 1 + \beta Y_2^0 (\theta))   ,
\end{equation}

\noindent 
where $\rho_N$ is the normalization constant, $\beta$ 
a quadrupole deformation parameter.  The 
radius parameter is determined as $ R = 1.2 A^{1/3}$ fm, 
and the diffuseness parameter is fixed to $a=0.5 $
 fm. For the small deformation, we expand the density in powers of 
$\beta$ and take first two terms as,

\begin{equation}
\bar{\rho}( {\bf r}) \approx \bar{\rho_0} (r) + \sqrt{ \frac{5}{16 \pi}} 
\bar{\rho_2} (r) Y_2^0 
(\theta)  ,
\end{equation}

\noindent 
where 

\begin{equation}
\bar{\rho_0} (r) = \frac{\rho_N}{1+{\rm exp\/}( \frac{r - R}{a})} \ \  ,
\end{equation}

\noindent 
and 

\begin{equation}
\bar{\rho_2} (r) = \sqrt{ \frac{16 \pi}{5} } \beta \rho_N \frac{R}{a} 
\frac{{\rm exp\/}( \frac{r - R}{a})}{ (1 + {\rm exp\/}( \frac{r - R}{a}))^2} \ \  .
\end{equation}

\noindent 
The nuclear densities in the laboratory frame $\rho (r)$ and $\rho_2 
(r)$, which should be used for 
our calculations for pionic atoms, are related to those in the intrinsic 
frame as follows,

\begin{displaymath}
\rho_0 (r) = \bar{\rho_0} (r)  , 
\end{displaymath}

\noindent
and

\begin{equation}
\rho_2 (r) = \frac{ J(2J-1) }{ (J+1) (2J+3) } \bar{\rho_2} (r) .  
\end{equation}

\noindent  
The $\beta$ parameter is related to the nuclear spectroscopic quadrupole 
moment $Q$ through, 

\begin{equation}
Q = \int_0^\infty \rho_2^p (r)  r^4 dr , 
\end{equation}

\noindent 
where a superscript $'p'$ indicates the proton density distribution.   
We can make similar approximation to the $\rho^2 (r)$ terms which appear 
in pion-nucleus optical potential as,

\begin{equation}
\rho^2 ( {\bf r}) \approx \rho_0^2 (r) + 2 \sqrt{ \frac{5}{16 \pi}} 
\rho_0 (r) \rho_2 (r) Y_2^0 (\theta) .
\end{equation}

\noindent
We can calculate pion wave functions by solving the Klein-Gordon 
equation with the Coulomb potential of a finite nuclear size, and with the 
pion-nucleus optical potential.  We include only spherical nuclear densities for  
the Klein-Gordon equation and evaluate the quadrupole deformation 
effect by perturbation.  We have applied the standard Ericson-Ericson 
potential \cite{Ericson66} with a parameter set determined by Seki 
$et$ $al.$ \cite{Seki83} and Konijn $et$ $al.$ \cite{Konijn90}.  
All the details of the calculation are described in 
Ref. \cite{Toki89}.  

Using the $\beta$ parameter, 
the potentials can be written as

\begin{displaymath}
V_{opt} = V_{opt}^{(0)} + \beta V_{opt}^{(2)}  ,
\end{displaymath}

\noindent 
and 

\begin{equation}
V_{C} = V_{C}^{(0)} + \beta V_{C}^{(2)}  .
\end{equation}

\noindent
The second terms are due to quadrupole deformation and treated as 
perturbation.  
Then, we can express the complex energy shift in the 
following way:

\begin{equation}
\Delta E = \beta  [ < V_{opt}^{(2)} > + < \frac{E-V_{C}^{(0)}}{\mu} 
V_{C}^{(2)} > ] , 
\end{equation}

\noindent 
where $E$ is the complex eigenvalue of the Klein-Gordon equation and 
$\mu$ the reduced mass of pion.  The terms coming from Coulomb and 
optical potential contribute to the complex energy shift differently as 
shown in Eq. (14) since these two potential appear in the 
Klein-Gordon equation in a different way as in Eq. (17).    
 
Because the Klein-Gordon equation includes the complex optical 
potential which makes the Hamiltonian non-Hermite and makes the 
eigen energies complex, we should be careful to the normalization 
condition to get proper orthonormal wave functions.  The 
Klein-Gordon equation can be rewritten as the following two-component 
equation \cite{Feshbach58};  

\begin{equation}
[ - (\sigma_3 + i \sigma_2) \frac{\mbox{\bf $ \nabla $}^2}{2 \mu}
+ \sigma_3 \mu + (\sigma_3 + i \sigma_2) V_{opt}^{(0)} + V_C^{(0)}] \Psi_i 
= E_i \Psi_i  ,
\end{equation}

\noindent 
where $\sigma_i$'s are the Pauli spin matrices and 

\begin{equation}
\Psi_i =  \frac{1}{2} \left(
         \begin{array}{c}
          (1 + (E-V_C^{(0)})/\mu) \phi_i \\
          (1 - (E-V_C^{(0)})/\mu) \phi_i
         \end{array}
         \right)   .
\end{equation}

\noindent 
This equation is equavalent to the Klein-Gordon equation, 

\begin{equation}
[ - \mbox{\bf $ \nabla $}^2 + \mu^2 + 2\mu V_{opt}^{(0)} ] \phi_i 
= (E_i-V_C^{(0)})^2 \phi_i .
\end{equation}

\noindent 
>From Eq. (15), we get orthonormal condition for $\Psi$ as;
\begin{equation}
\int \tilde \Psi ^T _i \sigma_3 \Psi_j d {\bf r} = \delta_{ij} ,
\end{equation}

\noindent 
where the $ \tilde \Psi $ means to take complex conjugate only for the 
angular part \cite{Feshbach58}.
This provides proper orthonormal condition for the 
wave functions as follows;

\begin{equation}
\int \tilde \phi_i \frac{E_i + E_j - 2V_{C}^{(0)}}{2 \mu} \phi_j d {\bf r} = 
\delta_{ij}
\end{equation}

\noindent
where the $ \tilde \phi $ means to take complex conjugate only for the 
angular part of the wave function.  
We have used the wave function, which satisfies Eqs.(17) and (19), 
to calculate all expectation values in 
this paper.  

Then, we evaluate the deformation effects on pionic atoms using 
the first 
order perturbation theory.  
To calculate the complex energy shift in Eq. (14), we need following expectation  
values for the deformation effect due to strong interaction: 

\begin{equation}
< \hat \rho ( {\bf r}) > \mid_{k=2} = - \frac{5}{8 \pi} C ( J l F ) 
\frac{l}{2l+3} \int_0^\infty R_{nl} (r) \rho_2 (r) R_{nl} (r) r^2 dr , 
\end{equation}

\begin{eqnarray}
< \mbox{\bf $ \nabla $} \hat \rho ({\bf r}) 
\mbox{\bf $ \nabla $} > \mid_{k=2} = &+&
\frac{5}{8 \pi} C ( J l F ) \frac{l}{2l+3} \{ \int_0^\infty 
\frac{dR_{nl}(r)}{dr} \rho_2(r) \frac{dR_{nl}(r)}{dr} r^2 dr 
\nonumber\\
&+& [l(l+1) - 3]  \int_0^\infty R_{nl}(r) \rho_2 (r) R_{nl}(r) dr 
\}, 
\end{eqnarray}

\begin{equation}
< \hat \rho ^2 ({\bf r})  > \mid_{k=2} = - \frac{5}{8 \pi} 
C ( J l F ) \frac{l}{2l+3} \int_0^\infty R_{nl}(r) (2 \rho_0 (r) \rho_2 (r) 
) R_{nl} (r) r^2 dr ,
\end{equation}

\noindent 
and 

\begin{eqnarray}
< \mbox{\bf $ \nabla $} \hat \rho ^2 ({\bf r}) 
\mbox{\bf $ \nabla $}> \mid_{k=2} = 
&+& \frac{5}{8 \pi} C ( J l F ) \frac{l}{2l+3} \{ \int_0^\infty 
\frac{dR_{nl}(r)}{dr} ( 2 \rho_0 (r) \rho_2 (r) ) \frac{dR_{nl}(r)}{dr} 
r^2 dr \nonumber\\
&+& [l(l+1) - 3]  \int_0^\infty R_{nl}(r) ( 2 \rho_0 (r) \rho_2 (r) ) 
R_{nl}(r) dr \}.  
\end{eqnarray}

\noindent 
For the deformation effects coming from Coulomb interaction, 
we need the matrix element: 

\begin{equation}
< \int \frac{\hat \rho^p ({\bf r'})}{\mid {\bf r} - {\bf r}' 
\mid} d^3 {\bf r}' >
\mid_{k=2} = - \frac{C ( J l F )}{2} \frac{l}{2l+3} 
\int_0^\infty r^2 dr \int_0^\infty r'^2 dr' \frac{r^2_<}{r^3_>} R_{nl}(r) 
R_{nl}(r) \rho^p_2 (r') .
\end{equation}

\noindent 
In above expressions, $R_{nl}(r)$ is the radial wave function of the 
bound pion and we have used the coefficient $C(J l F )$ defined as,

\begin{equation}
C (J l F) \equiv \frac{3X(X-1) - 4J(J+1)l(l+1)}{2J(2J-1)l(2l-1)}  ,
\end{equation}

\noindent
with

\begin{equation}
X \equiv J(J+1) + l(l+1) -F (F+1)  ,
\end{equation}

\noindent
where $l$ is the angular momentum of pionic states, and $F = l 
\otimes J$ the 
total angular momentum of the system. 
Obviously, the quadrupole deformation has no effect on pionic $s$ states 
within the first order perturbation.    \\

\noindent
{\bf 3. Numerical results}   \\

In this section, we evaluate the nuclear quadrupole effects on the deeply 
bound pionic atoms. 
We show the results obtained with the potential parameter set [a-1] in 
Table 1 of ref. \cite{Seki83}. 
In the numerical calculations, we have used the computer program MATOM 
\cite{Seki83} to obtain eigen energies and wave functions 
of pionic $2p$ and $3d$ states.
As a typical example of the deformed 
nucleus, we consider $^{175}_{71}$Lu$_{104}$ whose ground state is 
$J^{\pi}={7^+ \over  2}$ and quadrupole moment $Q=+3.49 ({\rm b})$.  
We fix the density parameters appeared in the previous section 
as $R=6.71 $ fm and 
$a=0.5 $ fm which give $\beta = 0.29$. By solving the Klein-Gordon 
equation with the spherical density, we obtain the binding energies 
and widths of the pionic $2p$ and $3d$ states which are 
B.E.($2p$)=4084, 
$\Gamma(2p)=258$ , B.E.($3d$)=2133 and $\Gamma($3d$)=29$ in unit of keV.  

First, we evaluate the quadrupole deformation effects on pionic $2p$ and $3d$ 
states.  In Fig. 1, we show 
the level shifts and widths of the pionic $2p$ state (a) and $3d$ 
state (b) as a function of the $\beta$ parameter.  
We can see in the Fig. 1 (a) that the $2p$ level splits into 
three levels with spacing of order $ 10 {\rm keV}$, which is one order of 
magnitude smaller than the width of the $2p$ state due to the nuclear 
absorption of bound pion.  Thus the quadrupole hyperfine splitting 
 will not change the level structure significantly 
for this state.  The $3d$ state, as shown in Fig. 2 (b), 
splits into five levels whose  
spacing is the same order of magnitude as the width of the $3d$ 
state.  In this case, the level splitting could be seen by proper experiments.   
For example, we could see the levels as small separated peaks on the top of the  
large peak consist of all $3d$ state contributions  
in double differential cross section of the (d,$^3$He) reactions, if we can 
achieve high energy-resolution like $ 10$  keV.  Even with worse energy 
resolution, we may be able to see the effects by observing the significantly 
larger width 
compared to those calculated with spherical nuclear density. 

Next, we evaluate the quadrupole effect for the nucleus with neutron 
skin.  This evaluation is much interesting since 
deeply bound pionic atoms are known to be sensitive to the neutron 
skin \cite{Toki90} and since the unstable nuclei are expected to have the 
neutron skin and the deformation simultaneously.
To simulate the existence of the neutron skin, we introduce a parameter $\Delta$  
as,  

\begin{equation}
\Delta = R_n - R_p ,
\end{equation}

\noindent
and vary the $\Delta$ by changing the radius of neutron distribution $R_n$.  
We show the calculated 
energies of deeply bound states as a function of $\Delta$ in Fig. 2.  
As shown in the figure, the deepest $1s$ state is most sensitive to the 
thickness of the neutron skin.  We also show the calculated energy 
levels of $2p$ and $3d$ atomic state as a function of deformation parameter 
$\beta$, in Fig. 
3 for $\Delta = 0.5 $ fm and in Fig. 4 for $\Delta = 1.0 $ fm.  
These figures show similar behavior as in Fig. 1 and indicate that 
the deformation effects on pionic states
are qualitatively same for nuclei with thick neutron skin.  

It is also interesting to compare the magnitude of the deformation 
effects with neutron skin effects.  For this purpose, we consider 
two cases 
$[\beta,\Delta ({\rm fm})] = [0.5, 0.0]$ and $[0.0, 1.0]$, and compare the 
calculated energies in each atomic state.  The results are followings: 

The deepest $1s$ state is very sensitive to the neutron skin.  The energy 
shift and width change due to neutron skin are 
$|\Delta E| \sim 300 $ keV and $|\Delta \Gamma| \sim 900 $ keV,  
while the energy of the $1s$ state is not affected by 
the deformation within the first order perturbation theory as mentioned 
at the end of section 2.  

The sensitivities of the $2p$ state to neutron skin 
are $|\Delta E| \sim 200 $ keV, 
$|\Delta \Gamma| \sim 10 $ keV. Those to deformation are 
$|\Delta E| = 10 \sim 40 $ keV, 
$|\Delta \Gamma| = 10 \sim 70 $ keV.  
The deformation effects depend on the total spin $F$ of the system 
defined in the last section. 

The sensitivities of the $3d$ state to neutron skin are 
$|\Delta E| \sim 20 $ keV, 
$|\Delta \Gamma| \sim 10 $ keV.  Those to deformation are 
$|\Delta E| = 0 \sim 20 $ keV, 
$|\Delta \Gamma| \le 10 $ keV.   

Before closing this section,  
we describe numerical results obtained with 
another set of potential parameter determined by Konijn $et$ $al.$
We used the parameter set 'present fit, $\xi =1$' shown in Table 1 of 
ref. \cite{Konijn90}.  
We have calculated the binding energies and widths of 2$p$ and 3$d$ state 
with the spherical nuclear density and obtained the similar results as 
those with Seki's potential \cite{Seki83}.  At $\Delta = 0$ [fm], we 
obtained B.E.$(2p)=4107$, $\Gamma (2p)=257$ and B.E.$(3d)=2140$, $\Gamma 
(3d) = 31$ in unit of keV.  

Then we evaluated the quadrupole deformation effects with Konijn's 
potential and compared them with those by Seki's [a-1] potential.  
For 3$d$ state, Konijn's potential gives 
40-60\% larger quadrupole shifts with 
the same sign.  As for the width, we obtained much larger deformation 
effects and its signs are opposite to those with Seki's potential at 
$ \Delta=0 $ and $ 0.5 $ fm, while both potentials give the quadrupole widths 
with the same sign for $ \Delta=1 $ fm.  For 2$p$ state, Konijn's potential 
gives opposite sign for both the real and the imaginary parts of the 
quadrupole energy shifts.  Concerning the absolute values, we obtain 
larger real energy shifts and smaller imaginary energy shifts with 
Konijn's potential for $ \Delta=0 $ and $ 0.5 $ fm, while for 
$ \Delta=1 $ fm Seki's potential yields larger real energy shifts 
and smaller imaginary energy shifts.  

We have found that the deformation effects are sensitive to the potential 
parameters because of the following reasons: As was pointed out by Seki 
and Masutani \cite{Seki83}, the potential parameters are correlated with 
each other.  The isoscalar s-wave term is approximately written as $(b_0 
+ B_0 \rho_e )\rho $ where $\rho_e$ is the effective nuclear density.  
Then the potential parameters with different $b_0$ and $B_0$ but with 
similar $(b_0 + B_0 \rho_e )$ value yield almost the same level shifts 
and widths for the pionic atom for spherical nuclei.  This is the same 
for p-wave terms.  In fact, the real part of the potential parameters by 
Seki [a-1] and Konijn [$\xi=1$] are quite differnt but the numerical 
values $(b_0 + B_0 \rho_e )$ are almost the same.  For the deformation 
effects, however, the different combination of the potential parameter 
appear as $<(b_0 + 2B_0 \rho )\rho_2> \sim (b_0+2B_0\rho_e) <\rho_2>$ and 
these are considerably different for the potentials by Seki [a-1] and 
Konijn.  Thus, the potentials which yield the similar results for the 
level shifts and widths of the spherical nuclei possibly give different 
quadrupole contribution.  Futhermore, there are large cancellation 
between the deformation effects coming from the strong and Coulomb 
interaction, so that the moderate changes of the strong quadrupole 
effects gives the large modification to the total deformation effects.  
Also, because of the complex pion wave function, the Coulomb term has 
contributions to real and imaginary energy shifts.  These are the reasons 
why the potentials by Seki and Konijn predicts fairly different 
quadrupole effects for deeply bound pionic atoms.  These situation does 
not happen for the quadrupole effects in shallow pionic states, because 
the Coulomb contribution dominates and there are no large cancellation 
between strong and Coulomb contributions.   \\

\noindent
{\bf 4. Summary}     \\

We have evaluated the nuclear quadrupole deformation effects on deeply 
bound pionic atoms.  The nuclear deformation induces the hyperfine 
splitting and affects the
widths of the pionic states, which are expected to be 
larger for deeper states since pion exists closer to the nucleus.  
We are very much interested in knowing the properties of the 
deeply bound pionic atoms for deformed nuclei and to find out 
the best atomic state - nucleus 
combinations for obtaining new information clearly.  

We found that the level shifts and width changes 
due to the quadrupole deformation are 
too small to observe for deeply bound $1s$ and $2p$ states.  
For $2p$ state the natural width of the state due to 
pion absorption is one order of magnitude larger than the level shift 
due to nuclear 
deformation. There are no quadrupole effects on pionic 1$s$ states 
in the first order perturbation.  
 Thus, it is very unlikely to observe the quadrupole 
effect in pionic $1s$ and $2p$ 
states.  For atomic $3d$ state, we have a chance to see the level splitting 
due to the deformation since the natural width of the $3d$ state is only 
around $30$  keV, which is comparable to the level shift due to the 
quadrupole deformation.  Thus, they may be observed with proper 
experimental technique like (d,$^3$He) reaction \cite{Hirenzaki91} 
with high 
energy-resolution.  

We have also studied the quadrupole effects for nuclei with neutron skin.  
This is interesting since we expect to have skin and deformation 
simultaneously in unstable nuclei, where we can produce deeply bound 
pionic states in principle using the (d,$^3$He) reaction in inverse 
kinematics \cite{Yamazaki91}.  
We change the radius of the neutron distribution artificially to simulate 
the neutron skin.  It is found that the existence of the neutron skin 
does not change the deformation effect significantly.  We compare the 
magnitudes of the both effects and summarize the results in section 3. 

We have evaluated the deformation effects using two different optical 
potential parameters and found that the nuclear quadrupole deformation 
effects are very sensitive to the potential parameters in deeply bound 
pionic states.  Thus, we can expect to determine the optical potential 
parameters more precisely if we can observe the nuclear deformation 
effects on deeply bound pionic atoms systematically.  

>From the results in this paper, we can conclude that the 
nuclear quadrupole deformation effects are difficult to observe 
experimentally for deeply bound pionic 
states like $1s$ and $2p$ orbits. In order to determine the deformation from the  pionic atoms, 
which could be 
interesting for unstable nuclei, we need to determine neutron skin thickness 
(monopole nuclear density) firstly by observing the $1s$ state and 
then determine the deformation from the observation of $3d$ state
with high energy-resolution like $10$ keV.

\vspace*{0.6cm}

\vspace*{0.6cm}

\noindent
{\bf Acknowledgements}

One of us (S. H.) acknowledges many discussions and 
collaborative works on the 
deeply bound pionic atoms with Prof. H. Toki, Prof. T. Yamazaki, 
and Prof. R. S. Hayano.

\pagebreak

\vspace*{0.6cm}

\noindent
{\bf Figure Caption} \\

\noindent
{\bf Fig. 1} \ \  Binding energies of (a) pionic $2p$ state and (b) $3d$ state 
as a function of the deformation parameter $\beta$ defined in the text.  
Total spin $F$ of each state is indicated in the figures.  
The width of each state is shown by the error bars.  \\ 

\noindent
{\bf Fig. 2} \ \  Binding energies of pionic $1s$, $2p$ and $3d$ states as a 
function of neutron skin thickness $\Delta$ defined in the text.  
The width of each state is shown as hatched areas.  The hatched 
area of $3d$ state is obtained by multiplying a factor 10 to the 
calculated width. \\ 

\noindent
{\bf Fig. 3} \ \ Same as in Fig. 1 except for neutron thickness parameter $\Delta 
= 0.5$ fm. \\  

\noindent
{\bf Fig. 4} \ \ Same as in Fig. 1 except for neutron thickness parameter 
$\Delta = 1.0$  fm.

\end{document}